\documentclass[pra,reprint,superscriptaddress,longbibliography]{revtex4-2}

\usepackage{array}
\usepackage{natbib}
\usepackage[section]{placeins}
\usepackage{pgfplots,amsfonts}
\usepackage[breaklinks]{hyperref}
\usepackage{amssymb}
\usepackage{amsmath,amsthm,bm}
\usepackage{amsfonts}%,dsfont}
\usepackage{cleveref}
\usepackage{verbatim}
\usepackage{mathtools}
\usepackage{tikz}
\hypersetup{citecolor=red,colorlinks=true,urlcolor=blue}
\usepackage{algorithm,setspace}
\usepackage{algpseudocode}
\usetikzlibrary{angles, quotes}
\usetikzlibrary{positioning}
\usepackage[normalem]{ulem}
\usepackage{color}
\usepackage{xcolor}

\newcolumntype{P}[1]{>{\centering\arraybackslash}p{#1}}

\newcommand{\bra}[1]{\mbox{$\langle #1 |$}}
\newcommand{\ket}[1]{\mbox{$| #1 \rangle$}}

\definecolor{Ugreen}{HTML}{198a11}

\begin{document}
\title{Shadow Ansatz for the Many-Fermion Wave Function in Scalable Molecular Simulations on Quantum Computing Devices}

\author{Yuchen Wang, Irma Avdic and David A. Mazziotti}

\email{damazz@uchicago.edu}

\affiliation{Department of Chemistry and The James Franck Institute, The University of Chicago, Chicago, Illinois 60637, USA}

\date{Submitted August 8, 2024}

\begin{abstract}
Here we show that shadow tomography can generate an efficient and exact ansatz for the many-fermion wave function on quantum devices. We derive the shadow ansatz---a product of transformations applied to the mean-field wave function---by exploiting a critical link between measurement and preparation. Each transformation is obtained by measuring a classical shadow of the residual of the contracted Schr{\"o}dinger equation (CSE), the many-electron Schr{\"o}dinger equation (SE) projected onto the space of two electrons. We show that the classical shadows of the CSE vanish if and only if the wave function satisfies the SE and, hence, that randomly sampling only the two-electron space yields an exact ansatz regardless of the total number of electrons. We demonstrate the ansatz's advantages for scalable simulations---fewer measurements and shallower circuits---by computing H$_{3}$ on simulators and a quantum device.
\end{abstract}

\maketitle

%The VQE routine can generally be divided into two steps, (\textit{i}) the state preparation, during which a parameterized circuit (ansatz) is used to represent a wave function on a quantum computer, and (\textit{ii}) the classical variational optimization of the circuit for expectation value calculations from the obtained measurements. 

\textit{Introduction.}---Simulating many-body systems is one of the most promising applications for near-term quantum computers~\cite{abrams1997, cao2019, Head-Marsden2020,mcardle2020, dutta2024}. Various algorithms that demonstrate a potential quantum advantage over classical ones have been proposed over time, including phase estimation~\cite{abrams1999, aspuru2005, du2010, paesani2017, Kitaev1995}, imaginary time evolution~\cite{mcardle2019, motta2020, kamakari2022, Tsuchimochi2023} and the variational quantum eigensolver (VQE)~\cite{Cao.2013, Peruzzo.2014, kandala2017, nakanishi2019, grimsley2019adaptive, tang2021}, with the VQE receiving significant attention due to its shallow circuit depth and adaptive ansatz form.  It is widely acknowledged that the chosen VQE ansatz has a profound impact on the performance characteristics of the algorithm, including circuit depth, convergence speed, and result accuracy~\cite{mitarai2019, grimsley2019adaptive, grimsley2019, sim2019, d2023}. For example, the chemistry-inspired unitary coupled cluster ansatz~\cite{bartlett1989, romero2018, anand2022, guo2024, windom2024, Mitra2024, lee2018, Hoffmann1988, Fedorov2022, Mehendale2023} suffers from a large parameter space~\cite{Wecker2015} while the hardware-efficient ansatz~\cite{kandala2017, d2023} tailors the parametrization to the quantum device but is limited by the barren plateau phenomenon~\cite{Wiersema2020, Cerezo2021, Park2024}. 

In this Letter, we derive an efficient and exact ansatz for the many-fermion wave function on quantum devices by combining concepts from shadow tomography~\cite{Aaronson2020, Huang2020, huang2022, elben2023, zhao2021, nguyen2022, hu2023, akhtar2023, wan2023, ippoliti2023, helsen2023, bertoni2024, wu2024, bu2024, vermersch2024, avdic2024, Hearth.2023, Gyurik.2023, Jerbi.2023, Lewis.2024, O'Gorman.2022, Koh.2022, Coopmans.2023, Guzman.2023, Jnane.2024, Hu.2024, Truger.2024, Caprotti.2024, Majsak.2024, Levy.2024, Becker.2024, Somma2024} and reduced density matrix theory~\cite{Mazziotti.2007, Coleman.2000, Mazziotti.2012, Piris.2021, Mazziotti.2023}. We obtain each transformation in the shadow ansatz---a product of transformations applied to the mean-field reference---by measuring a classical shadow of the residual of the contracted Schr{\"o}dinger equation (CSE)~\cite{Mazziotti1998, Nakatsuji1996, Colmenero1993, valdemoro1997, mazziotti1999, mazziotti2002}, the many-electron Schr{\"o}dinger equation (SE) projected onto the space of two electrons.  Selecting each classical shadow with respect to a randomly sampled one-electron basis, we show, completely covers the quantum two-electron space with respect to both measurement and preparation. Moreover, we prove that the classical shadows of the CSE vanish if and only if the wave function satisfies the SE and, hence, the shadow ansatz is exact regardless of the quantum system's total number of electrons. Advantages of the shadow ansatz, relative to solving the CSE or its anti-Hermitian part with full tomography~\cite{Smart2021, smart2022benzyne, smart2022encoding, Smart.2022, smart2024, wang2023, wang2023boson, benavides2024, warren2024}, include fewer measurements and shallower circuits.  

We first present the theory of the shadow ansatz, followed by quantum simulation results for the linear H$_3$ chain using a noiseless simulator and a noisy 127-qubit IBM quantum computer. We examine the differences and advantages of the shadow ansatz compared to the existing ansatz and showcase its ability to reduce computational costs for scalable simulation on near-term quantum devices.

\textit{Theory.---}
%The outline for theory
%1.CSE and SE, mention the equivalence
%2.measure of the residual yields the state preparation, conventional ansatz 
%3.shadow tomography comes into play, introduce
%4.shadow ansatz form (shadow state preparation)
%5.now migrate to explain ACSE
%6.mention the measurement technique imaginary time evolution to compute the commutator in ACSE
%move the connection and comparison to other ansatz at the end of the paper with (discussion and conclusion)
Let us first briefly recall CSE theory. Consider the SE for an $N$-electron system
\begin{equation}
    (\hat{H}-E)\ket{\Psi} = 0,
\end{equation}
where $\hat{H}$ is the Hamiltonian operator and $\ket{\Psi}$ is the wave function. Projecting the SE onto the two-electron space yields the CSE~\cite{Mazziotti1998, Nakatsuji1996, Colmenero1993, valdemoro1997, mazziotti1999, mazziotti2002} 
\begin{equation}\label{eq:cse}
    \bra{\Psi}\hat{a}^{\dagger}_i\hat{a}^{\dagger}_j\hat{a}^{}_l\hat{a}^{}_k(\hat{H}-E)\ket{\Psi} = 0,
\end{equation}
where $\hat{a}_{i}^{\dagger}$ and $\hat{a}^{}_{i}$ are the fermionic creation and annihilation operators for orbital $i$. While the SE clearly implies CSE, we can prove the reverse that the CSE implies the variance that implies the SE~\cite{Mazziotti1998, nakatsuji1976}, demonstrating that the CSE and SE share a common set of energetically non-degenerate pure-state solutions. 

Iterative solution of the CSE suggests an exact product ansatz for the wave function~\cite{Mazziotti.2004,Mazziotti.2020,smart2024}
\begin{equation}
    | \Psi_{n+1} \rangle = \prod_{q=0}^{n}{e^{-\eta_{q}\hat{R}_{q}^{\dagger}} | \Psi_0 \rangle}, \label{eq:ansatz}
\end{equation}
where
\begin{equation}
    \hat{R}_q = \sum_{ijkl}{ {^{2} R^{ij;kl}_{q}}\hat{a}^{\dagger}_i\hat{a}^{\dagger}_j\hat{a}^{}_l\hat{a}^{}_k }.
\end{equation}
Descent along the gradient of the energy with respect to the wave function's parameters $^{2} R^{ij;kl}_{n}$ at the $n^{\rm th}$ iteration ($q=n$) occurs when these parameters equal the residual of the CSE.  Hence, the gradient at a given iteration vanishes if and only if the CSE and thus, the SE are satisfied, proving the ansatz's exactness~\cite{Mazziotti.2004}.

%in which the matrix elements ${^{2} R^{ij;kl}_{n}}$ represent the residual of the CSE with respect to the approximate wave function at the $n^{\rm th}$ iteration. Significantly, at the $n^{\rm th}$ iteration the gradient of the energy with respect to the elements of ${^{2} R^{ij;kl}_{n}}$ in the above ansatz---the CSE ansatz---vanishes around ${^{2} R^{ij;kl}_{n}} = 0$ if and only if the CSE vanishes, implying that the SE is satisfied and proving the ansatz's exactness. The ansatz establishes a significant connection between state preparation and measurement by using the measured residual $\hat{R}$ at the $n^{\rm th}$ iteration to update the wave functions for the next iteration.

%The CSE offers a unique solution to simulate many-electron systems on quantum computers from the two-particle reduced density matrix (2-RDM) tomography. We observe that the exponential form of the CSE residual in Eq. \ref{eq:cse} can be used to converge the CSE. The corresponding ansatz reads,
%\begin{equation}
%    | \Psi_{n} \rangle = \prod_{n}e^{\eta_{n}\hat{R}_{n}} | \Psi_0 \rangle,
%\end{equation}
%$\eta$ is the learning rate and the residual $\hat{R}$ can be estimated with partial RDM tomography on quantum computers. 

Here we exploit this connection by drawing inspiration from the theory of shadow tomography~\cite{Aaronson2020, Huang2020, huang2022, elben2023, zhao2021, nguyen2022, hu2023, akhtar2023, wan2023, ippoliti2023, helsen2023, bertoni2024, wu2024, bu2024, vermersch2024, avdic2024, Hearth.2023, Gyurik.2023, Jerbi.2023, Lewis.2024, O'Gorman.2022, Koh.2022, Coopmans.2023, Guzman.2023, Jnane.2024, Hu.2024, Truger.2024, Caprotti.2024, Majsak.2024, Levy.2024, Becker.2024, Somma2024} to prepare a shadow ansatz, which enhances both state preparation and measurement with reduced circuit depth and measurement cost. Rather than directly projecting the SE onto the two-electron space, we project the SE onto a set of classical shadows $^{2} S_{q}^{ij}$ where each classical shadow is the diagonal (or classical) part of the two-electron space with respect to a given random unitary transformation $\hat{U}_{q}$ of the orbitals
\begin{equation}
    ^2S_{q}^{ij} = \bra{\Psi}\hat{U}^{\dagger}_{q}\hat{a}^\dagger_{i}\hat{a}^\dagger_{j}\hat{a}^{}_{j}\hat{a}^{}_{i} \hat{U}^{}_{q} (\hat{H}-E)\ket{\Psi} .
    \label{eq:shadow} 
\end{equation}
Each classical shadow of dimension $r^{2}$ where $r$ is the number of spin orbitals is a subset of the CSE residual of dimension $r^{4}$. Because there are $r^{2}$ linearly independent unitary transformations of the orbitals, connecting each of the diagonal orbitals to each of the off-diagonal orbitals, a set of $O(r^{2})$ classical shadows of total dimension $r^{4}$ represents a shadow projection of the SE onto the two-electron space---a shadow CSE---that is equivalent to the CSE.  Consequently, we have the following generalization of the exactness of the CSE: a wave function satisfies the shadow CSE if and only if it satisfies the SE.

Building upon the parallels between the shadow CSE and the CSE, we can define a shadow ansatz for the many-fermion wave function 
\begin{equation}
    | \Psi_{n+1} \rangle = \prod_{q=0}^{n}{ \left ( \prod_{t=1}^{M}{e^{-\eta_{qt}\hat{S}_{qt}^{\dagger}}} \right )} | \Psi_0 \rangle , \label{eq:sansatz2}
\end{equation}
where
\begin{equation}
    \hat{S}_{qt} = {\hat U}^{}_{qt} \left ( \sum_{ij}{ {^{2} S^{ij}_{qt}}\hat{a}^{\dagger}_i\hat{a}^{\dagger}_j\hat{a}^{}_j\hat{a}^{}_i } \right ) {\hat U}^{\dagger}_{qt}. \label{eq:S_hat2}
\end{equation}
Each of the $M$ classical shadows in the $n^{\rm th}$ iteration is defined with respect to the $n^{\rm th}$ wave function, using a different random unitary transformation.  In the limit that $M$ approaches $r^{2}$, the shadow ansatz effectively becomes the previously developed CSE ansatz in Eq.~(\ref{eq:ansatz}). The value of $M$ can also be varied between iterations, allowing one to tune between the CSE ansatz and the single-shadow-per-iteration ($M=1$) ansatz.  Like the CSE ansatz, we can follow the energy gradient descent direction by setting the parameters at the $n^{\rm th}$ iteration to the $M$ classical shadows of the CSE residual, as shown for one classical shadow in Eq.~(\ref{eq:shadow}).  Hence, the energy gradient at the $n^{\rm th}$ iteration vanishes if and only if the $M$ classical shadows of the residual vanish.  The shadow ansatz becomes exact if we continue its construction until $O(r^{2})$ classical shadows of the CSE vanish, which implies the CSE and hence, the SE.

To implement the ansatz on quantum simulators or devices, we further split the classical shadow of the residual in Eq.~(\ref{eq:shadow}) into its Hermitian and anti-Hermitian parts, displayed below as the anti-commutator and commutator respectively,
\begin{equation}\label{eq:hcse+acse}
\begin{split}
    2 \; ^2S_{qt}^{ij} = \bra{\Psi_{q}}\{\hat{U}_{qt}^{\dagger}\hat{a}^\dagger_{i}\hat{a}^\dagger_{j}\hat{a}^{}_{j}\hat{a}^{}_{i} \hat{U}_{qt}^{} (\hat{H}-E)\}\ket{\Psi_{q}} \\ +\bra{\Psi_{q}} [\hat{U}_{qt}^{\dagger}\hat{a}^\dagger_{i}\hat{a}^\dagger_{j}\hat{a}^{}_{j}\hat{a}^{}_{i} \hat{U}_{qt}^{},\hat{H}]\ket{\Psi_{q}}.
\end{split}
\end{equation}
The anti-Hermitian portion, known as the anti-Hermitian CSE (ACSE)~\cite{Mazziotti2006, mazziotti2007acse, Mazziotti.2007b, Rothman.2009, gidofalvi2009, Snyder.2011, Boyn.2021}, can be readily implemented via unitary transformations on quantum computers~\cite{smart2020, smart2022encoding}. While ACSE does not strictly imply the CSE~\cite{Mazziotti.2004}, practical calculations on quantum simulators and devices show that, at least for the molecular systems explored, it converges to the energies and 2-RDMs from full configuration interaction (FCI)~\cite{Smart2021, smart2022benzyne, smart2022encoding, Smart.2022, smart2024, wang2023, wang2023boson, benavides2024, warren2024}. We note that, as shown in recent work~\cite{smart2024}, it is possible to implement the full residual and non-unitary transformation on quantum computers with dilation techniques~\cite{Hu.2020, Schlimgen.2021}. 

\begin{figure}
    \centering
    \includegraphics[width=0.5\textwidth]{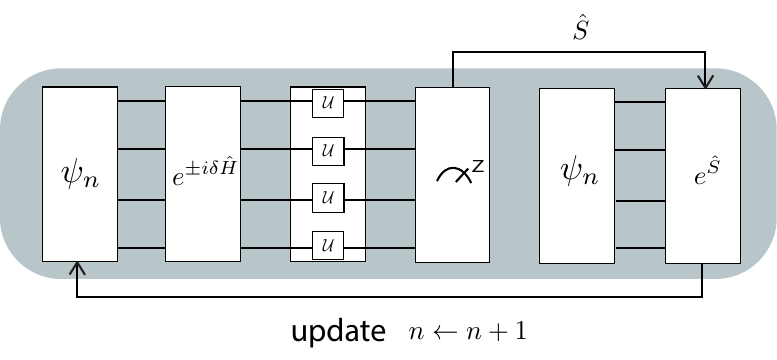}
    \caption{A schematic representation of the CQE algorithm with the shadow ansatz.}
    \label{fig:flowchart}
\end{figure}

To obtain the classical shadow of the ACSE part of the residual in Eq.~(\ref{eq:hcse+acse}), we measure
\begin{equation}
\begin{split}\label{eq:commutator}
    ^{2}S_{qt}^{ij} = \frac{1}{2i\delta}(\bra{\Lambda^{+}_{q}}\hat{U}^{\dagger}_{qt}\hat{a}^\dagger_{i}\hat{a}^\dagger_{j}\hat{a}^{}_{j}\hat{a}^{}_{i}\hat{U}_{qt}\ket{\Lambda^{+}_{q}} \\ -\bra{\Lambda^{-}_{q}}\hat{U}^{\dagger}_{qt}\hat{a}^\dagger_{i}\hat{a}^\dagger_{j}\hat{a}^{}_{j}\hat{a}^{}_{i}\hat{U}_{qt}\ket{\Lambda^{-}_{q}})
    + O(\delta^{2}),
\end{split}
\end{equation}
where $\ket{\Lambda^{\pm}_{q}}=e^{\pm i \delta H}\ket{\Psi_{q}}$ in which $\delta$ is the stepsize~\cite{smart2020}. Table~\ref{table1} and Fig.~\ref{fig:flowchart} summarize the iterative algorithm---a contracted quantum eigensolver (CQE)---that solves the ACSE with the shadow ansatz. In the results presented here, we form the unitary transformations $\hat{U}$ from the tensor product of random single-qubit unitaries drawn from the single-qubit Clifford subgroup. Other choices for the unitary ensemble can be readily incorporated into the algorithm for future studies and potential improvements~\cite{zhao2021, nguyen2022, hu2023, helsen2023}. 

%\begin{figure}[h]
%        \begin{algorithm}[H]
%        \begin{spacing}{1}
        %\setstretch{1.09}
%        \begin{algorithmic}[1]
%        \State prepare initial state $\Psi^{(0)}$,
%        \State Set $n \gets 0$,
%        \While{$||\hat{R}_{n}|| > \delta$}
%            \LState choose number of shadows $m$
%            \LState Set $R \gets 0$ 
%            \For {$0\leq \nu < m$}
%            \LState prepare $\ket{\lambda^\pm} =  U_{\nu}e^{\pm i \eta \hat H} \ket{\Phi^{(n)}}$,
%            \LState $R_{n} \gets R_{n} +  \frac{1}{2\eta i}\sum_{z=\pm} z \bra{\lambda^z} {U^{\dagger}_{\nu}\hat \Gamma}^{pq}_{qp}U_{\nu}\ket{\lambda^z}$,
%            \EndFor
%            \LState take $\theta^* = {\rm argmin} \bra{\Psi^{(n)}(\theta)}\hat H\ket{\Psi^{(n)}(\theta)}$,
%            \LState prepare $\ket{\Psi^{(n+1)}(\theta)} = e^{\theta \hat A^{(n)}} \ket{\Psi^{(n)}}$,
%            \LState $n \gets n+1$.
%        \EndWhile
%        \end{algorithmic}
%        \caption{CQE with shadow ansatz}
%        \label{alg:cap2}
%        \end{spacing}
%        \end{algorithm}
%\end{figure}

\begin{table}[t!]
  \caption{\normalsize CQE algorithm that solves the ACSE with the shadow ansatz.}
  \label{table1}
  \begin{ruledtabular}
  \begin{tabular}{l}
  {\bf Algorithm: CQE with the shadow ansatz} \\
  \hspace{0.0in} {\rm Given} $n=0$ {\rm and convergence tolerance} $\epsilon$. \\
  \hspace{0.0in} {\rm Choose initial wave function} $| \Psi_{0} \rangle$. \\
  \hspace{0.0in} {\rm Repeat until the residual is less than} $\epsilon$: \\
  \hspace{0.1in} {\rm {\bf Step 1:} Prepare $\ket{\Lambda^{\pm}}=e^{\pm i \delta H}\ket{\Psi_n}$.}   \\
  \hspace{0.1in} {\rm {\bf Step 2:} Measure $M$ shadows with Eq.~(\ref{eq:commutator}).} \\
  \hspace{0.1in} {\rm {\bf Step 3:} Update $\ket{\Psi_n}$ from $M$ shadows.} \\
  \hspace{0.1in} {\rm {\bf Step 4:} Optimize energy with respect to $\eta_{nt}$.} \\
  \hspace{0.1in} {\rm {\bf Step 5:} Set} $n \leftarrow n+1$.
  \end{tabular}
  \end{ruledtabular}
\end{table}

The number of Pauli exponential terms in the proposed algorithm scales as $O(\alpha r^{2})$, where $\alpha$ is the number of repetitive measurements and $r$ is the number of orbitals, in which $\alpha$ has been shown to scale polylogarithmically with the system size~\cite{Aaronson2020}. Compared to the full 2-RDM tomography with an $O(r^{4})$ cost, the shadow ansatz is advantageous in the number of measurements, and, even more importantly, since the ansatz directly utilizes the measurement outcomes, the improvement over conventional tomography is also manifested in the circuit depth.

\emph{Results}---We use the CQE with the shadow ansatz to perform calculations on the linear molecule H$_{3}$ with equally spaced bond lengths on a noiseless six-qubit statevector simulator and the noisy 127-qubit IBM Cleveland device~\cite{ibm_quantum, Qiskit}. We compute the spin sector $\langle {\hat S}_z \rangle=1$ for H$_{3}$ with the Slater-type Gaussian orbital (STO-3G) basis set~\cite{Hehre1969}. Electron integrals and FCI are computed with PySCF~\cite{Sun.2018}.  For the noiseless simulator, we use the Jordan-Wigner mapping~\cite{Jordan1928, Fradkin1989} to map the six spin-orbitals to six qubits. For IBM Cleveland, additional tapering techniques are used to map the Hamiltonian to three qubits. More details on the implementation can be found in the Supplemental Material (SM).

\begin{figure}[th!]
  \centering
  \includegraphics[width=0.45\textwidth]{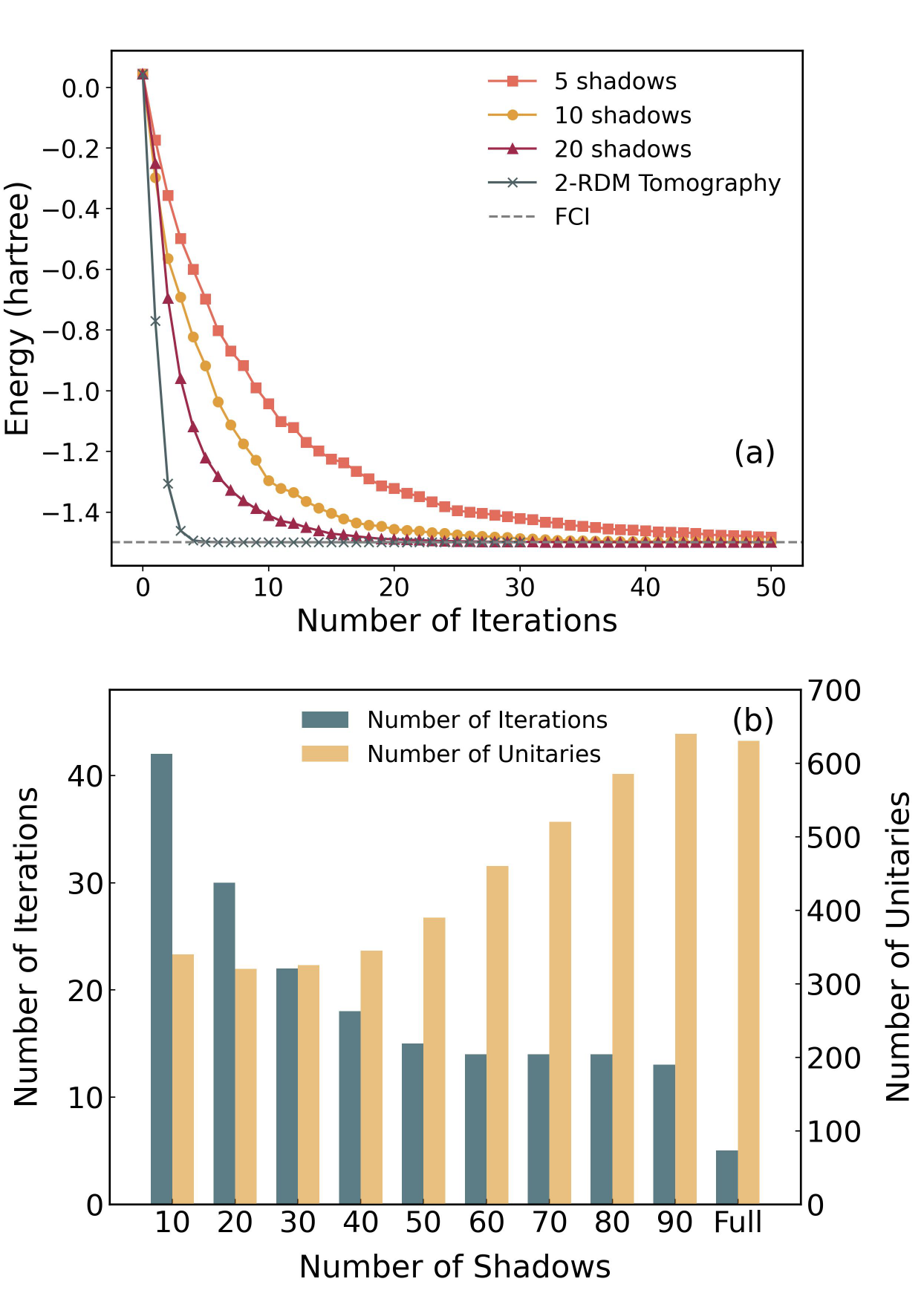}
  \caption{Convergence data for the linear molecule H$_{3}$ with equal bond lengths of 0.7~\AA. (a) As a function of the iteration number, the energies from the shadow ansatz with different numbers $M$ of shadows are compared with the energy from the CSE ansatz with conventional 2-RDM tomography. (b) As a function of the number $M$ of shadows per iteration, the number of iterations and the circuit depth, required to reach energy convergence within 1~mhartree, are compared.}
  \label{fig:1}
\end{figure}

Figure~\ref{fig:1} (a) illustrates the convergence of the CQE ground-state energy of H$_{3}$ to the FCI~\cite{Craig1950} wave function solution, as a function of the number $M$ of shadows per iteration, on a noiseless simulator. The initial trial wave function is the normalized and unweighted sum of all possible spin-adapted Slater determinants to minimize the potential effect of the initial guess. All calculations show exact convergence to the FCI solution, even with the imposed randomized measurement. This shows that the shadow ACSE ansatz with as few as five shadows ($M=5$) may be sufficient to converge rapidly to the FCI solution. The more shadows employed, the fewer iterations are needed for the convergence, which can be viewed as the shadow residual approaching the exact residual limit.

In Fig.~\ref{fig:1} (b), we further compare the circuit depth of the shadow ansatz with different numbers $M$ of shadows per iteration. The circuit depth is represented by the number of elementary Pauli exponential terms after first-order Trotterization. For a number of shadows fewer than 40,  as more shadows are employed, the number of iterations decreases but the circuit depth remains approximately constant, suggesting a trade-off between the number of iterations and the number $M$ of shadows per iteration. We also observe that once the number $M$ of shadows per iteration exceeds 40, further increasing $M$ does not significantly accelerate the convergence, but instead increases the circuit depth, leading to two important conclusions. First, the number $M$ of shadows employed to minimize the residual to a desired level of accuracy should not exceed an upper limit to maximize circuit efficiency. Second, provided that the number of shadows remains below the established upper limit, the distribution of shadows across iterations has a rather small impact on circuit depth. This flexibility allows for the dynamic manipulation of the shadow ansatz. It is also worth noting that the exact 2-RDM tomography leads to the fewest iterations but not the optimal circuit depth, which can be attributed to the fact that the residual from exact tomography usually contains many trivial terms that do not contribute significantly to the optimization direction.

\begin{figure}[t!]
  \centering
  \includegraphics[width=0.4\textwidth]{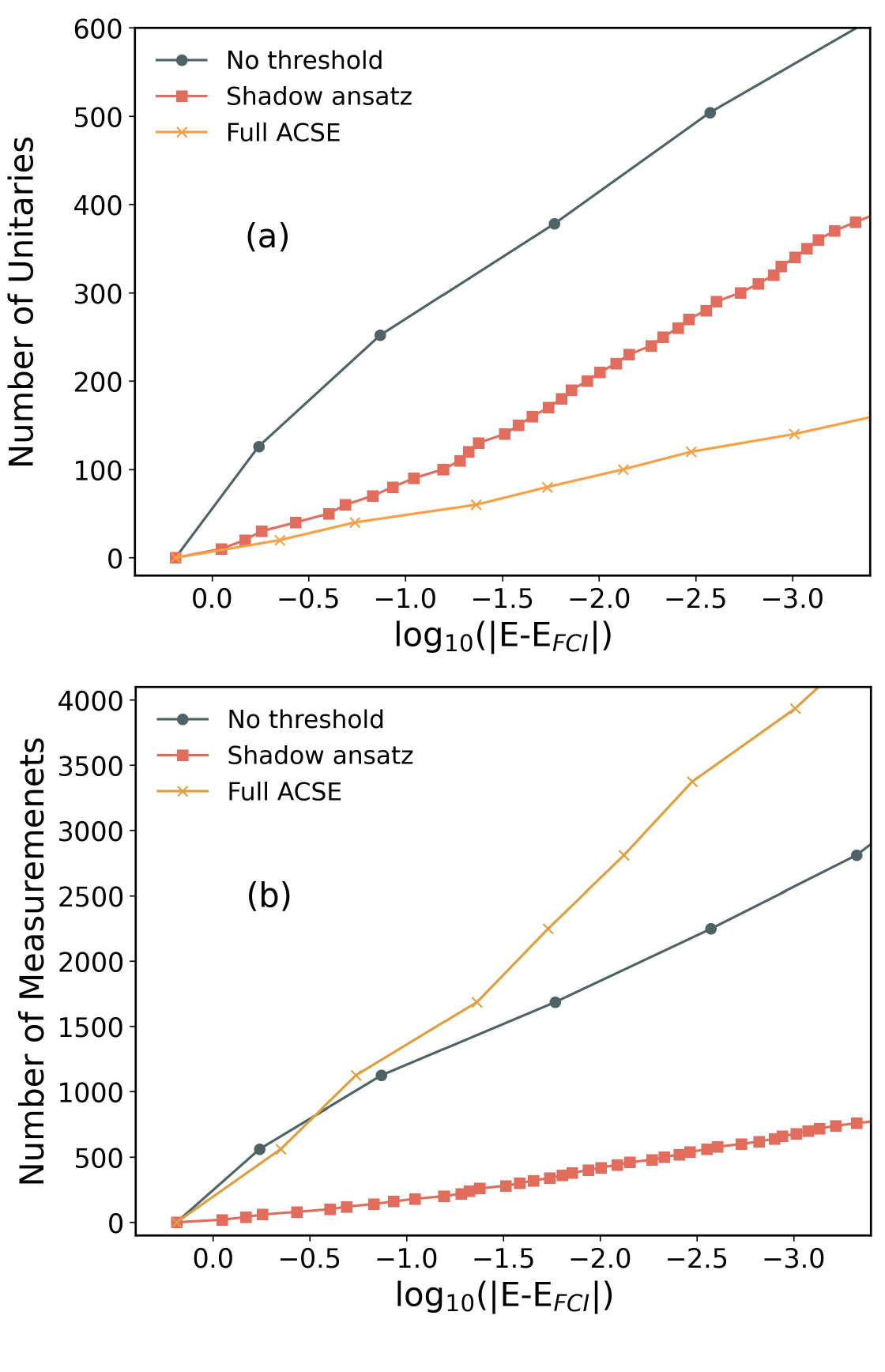}
  \caption{(a) The number of accumulated unitaries and (b) the number of required measurements plotted against accuracy during optimization. The energy obtained from the full configuration interaction (or exact diagonalization) is denoted as E$_{\text{FCI}}$. 20 shadows are used per iteration. The number of measurements refers to the number of distinct circuits, which should be distinguished from the number of shots applied to each circuit reported in the SM.}
  \label{fig:3}
\end{figure}

Due to the decoherence effect prevalent on current quantum computers, it is necessary to maintain a low circuit depth for any practical algorithm. In previous work, we employed a threshold to filter small contributions from the residual. A straightforward implementation of this thresholding is to decompose the residual in the Pauli basis and filter the coefficients of the terms based on their magnitude, which creates a linear increase in circuit depth with respect to the number of iterations. The effect of thresholding is illustrated in Fig.~\ref{fig:3} (a). Both the shadow ansatz and the conventional ACSE ansatz, after applying the threshold, achieve a significant reduction in circuit depth, measured in the number of unitaries, compared to the results with no applied thresholding. The conventional ACSE ansatz provides slightly better convergence with the number of unitaries than the shadow ansatz. However, concerning the number of measured circuits, shown in Fig.~\ref{fig:3} (b), the shadow ansatz significantly outperforms the conventional ACSE ansatz. The threshold in conventional ACSE ansatz is performed in a post-selection of unitaries with designated measurement instructions, which means that we still measure much more information than needed. In contrast, the shadow ansatz provides a protocol for thresholding in an ``on-the-fly'' manner with random sampling decreasing the number of measurements.  

%In contrast, the full tomography, even with thresholding, does not tailor the measurement we need to perform to reflect its ultimate importance in the ansatz.

\begin{figure}[t]
  \centering
  \includegraphics[width=0.45\textwidth]{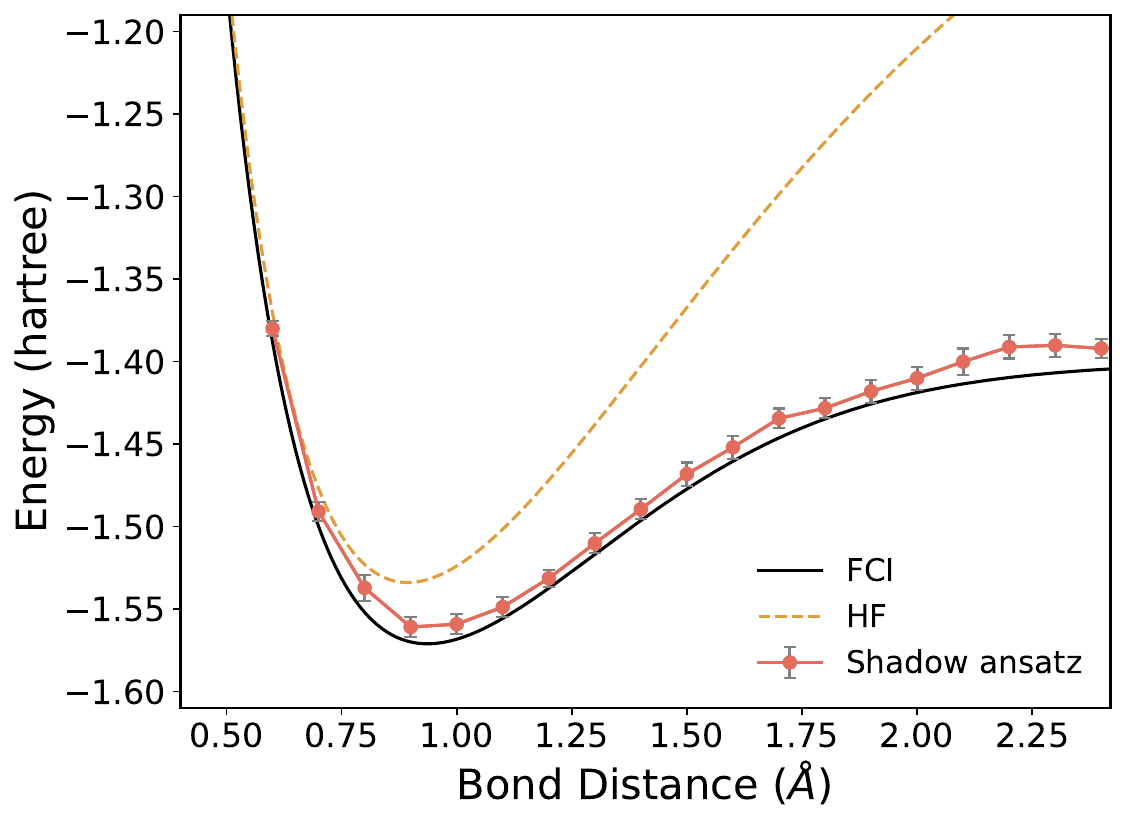}
  \caption{The dissociation curve of linear H$_{3}$ obtained with Hartree-Fock (HF), full configuration interaction (FCI), and the shadow ansatz. The hydrogen atoms are kept equally spaced during bond stretching. The data point and error bar are plotted as mean value $\pm$ standard deviation.}
  \label{fig:4}
\end{figure}

We examine the performance of the shadow ansatz on realistic quantum computers by performing the simulation of linear H$_{3}$  on the 127-qubit IBM Cleveland device~\cite{ibm_quantum}. The dissociation curves of H$_{3}$, obtained from the Hartree-Fock (HF), full configuration interaction (FCI), and the shadow-ansatz methods are plotted in Fig.~\ref{fig:4}.  Bond lengths are kept equally spaced upon dissociation.  The curve from the initial HF trial wave function is indicated by the dashed lines. The shadow ansatz is able to capture the correlation energy from the HF guess, especially in the dissociated region where HF does not perform well due to its restriction to a single Slater determinant. The results obtained from shadow ansatz differ from FCI on average by approximately 16~mhartree, which is quite accurate given the degree of noise on the near intermediate-scale quantum (NISQ) computers. 

%More accurate result could be obtained with advanced error mitigation techniques yet will not be pursued here.

\textit{Discussion and conclusions}---
Two challenges for the quantum simulation of many-electron quantum systems are: ({\em i}) measurement of observables and ({\em ii}) preparation of the wave function.  Classical shadows---classical snapshots that collectively represent a quantum system---were developed to improve the efficiency of measurements in ({\em i}).  Here we combine shadow tomography~\cite{Aaronson2020, Huang2020, huang2022, elben2023, zhao2021, nguyen2022, hu2023, akhtar2023, wan2023, ippoliti2023, helsen2023, bertoni2024, wu2024, bu2024, vermersch2024, avdic2024, Hearth.2023, Gyurik.2023, Jerbi.2023, Lewis.2024, O'Gorman.2022, Koh.2022, Coopmans.2023, Guzman.2023, Jnane.2024, Hu.2024, Truger.2024, Caprotti.2024, Majsak.2024, Levy.2024, Becker.2024, Somma2024} with 2-RDM theory~\cite{Mazziotti.2007, Coleman.2000, Mazziotti.2012, Piris.2021, Mazziotti.2023} to improve the preparation in ({\em ii}).   Measuring the two-electron CSE---the projection of the many-electron SE onto the two-electron space---provides all the information necessary to construct an exact ansatz for the many-electron wave function~\cite{Mazziotti.2004}.  The CSE from 2-RDM theory, therefore, provides a critical link between measurement and preparation.  By combining the CSE with shadow tomography, we obtain a highly efficient exact ansatz---the shadow ansatz---in which the many-electron wave function is directly constructed from the classical shadows of the CSE.  Relative to full tomography, the shadow ansatz provides both a decrease in the circuit depth (when compared to the ACSE ansatz without thresholding) and a significant reduction in the number of measurements needed to construct the wave function.  The improvements in efficiency have important implications for realizing scalable molecular simulations.  Future work will explore further extensions and improvements of the shadow ansatz including the use of optimal shadow tomography protocols beyond those implemented here.       
%We first briefly comment on the difference between the CSE ansatz and other popular ansatze and then discuss the improvement of the shadow ansatz over the CSE ansatz. Unlike the truncated coupled-cluster ansatz and its unitary variant, which contain only occupied-to-virtual excitation, the CSE ansatz is exact~\cite{Mazziotti.2004, Mazziotti.2020, smart2024}.  Generalized coupled-cluster ansatz~\cite{lee2018} overcomes the occupied-to-virtual issue, but by restricting the two-body operator in the CSE to be anti-Hermitian, we obtain the ACSE ansatz which is the complete set of two-body unitary transformation one can extract from the many-electron Schr\"{o}dinger equation~\cite{Mazziotti.2007, smart2020}. The shadow ansatz built from CSE inherits the advantages above and further improves the efficiency. The improvement is reflected in both a decrease in the circuit depth (at least when compared to the ACSE ansatz without thresholding) and most importantly, a reduction in the number of measurements to construct the wave function.  The shadow ansatz combines the advantages of classical shadows for tomography with reduced density matrix theory and in particular, the CSE, which provides a critical connection between preparation and measurement, to improve the accuracy and efficiency of wave function preparation on quantum devices.  Future work will explore further improvements of the shadow ansatz from optimal shadow tomography protocols beyond the implementation in this work.

\textit{Acknowledgments}---D.A.M gratefully acknowledges the U.S. National Science Foundation Grant No. CHE-2155082 and the U.S. Department of Energy, Office of Basic Energy Sciences, Grant DE-SC0019215. I.A. gratefully acknowledges the NSF Graduate Research Fellowship Program under Grant No. 2140001. The views expressed are of the authors and do not reflect the official policy or position of IBM or the IBMQ team.

\bibliography{Refs2,main}
\end{document}